\documentclass[journal=nalefd,manuscript=article]{achemso}
\usepackage[version=3]{mhchem} 
\usepackage{graphicx}

\usepackage{amsfonts}
\usepackage{amsmath}
\usepackage{amssymb}
\usepackage{amsbsy}
\usepackage{setspace}
\usepackage{xcolor}
\usepackage{ulem}

\usepackage{pdfpages}

\newcommand*\arucl{{$\alpha$-RuCl$_3$}}

\author{Jesse Balgley}
\author{Jackson Butler}
\affiliation{Department of Physics, Washington University in St.\ Louis, 1 Brookings Dr., St.\ Louis MO 63130, USA}

\author{Sananda Biswas}
\affiliation{Institut f\"ur Theoretische Physik, Goethe-Universit\"at Frankfurt, 60438 Frankfurt am Main, Germany}

\author{Zhehao Ge}
\affiliation{Physics Department, UC Santa Cruz, 1156 High Street, Santa Cruz, CA 95064, USA}

\author{Samuel Lagasse}
\affiliation{Electronics Science and Technology Division, United States Naval Research Laboratory, Washington, DC 20375, United States}

\author{Takashi Taniguchi}
\affiliation{International Center for Materials Nanoarchitectonics, National Institute for Materials Science, 1-1 Namiki, Tsukuba, 305-0044, Japan}

\author{Kenji Watanabe}
\affiliation{Research Center for Functional Materials, National Institute for Materials Science, 1-1 Namiki, Tsukuba, 305-0044, Japan}

\author{Matthew Cothrine}
\affiliation{Material Science \& Technology Division, Oak Ridge National Laboratory, Oak Ridge, Tennessee 37831, USA}

\author{David G.\ Mandrus}
\affiliation{Material Science \& Technology Division, Oak Ridge National Laboratory, Oak Ridge, Tennessee 37831, USA}
\affiliation{Department of Material Science and Engineering, University of Tennessee, Knoxville, Tennessee 37996, USA}

\author{Jairo Velasco Jr.}
\affiliation{Physics Department, UC Santa Cruz, 1156 High Street, Santa Cruz, CA 95064, USA}

\author{Roser Valent\'i}
\affiliation{Institut f\"ur Theoretische Physik, Goethe-Universit\"at Frankfurt, 60438 Frankfurt am Main, Germany}

\author{Erik A.\ Henriksen}
\affiliation{Department of Physics, Washington University in St.\ Louis, 1 Brookings Dr., St.\ Louis MO 63130, USA}
\email{henriksen@wustl.edu}

\title{Ultra-sharp lateral $p\text{-}n$ junctions in modulation-doped graphene}


\begin{document}

\date{\today}

\begin{abstract}
We demonstrate ultra-sharp (${\lesssim}\,10\text{ nm}$) lateral $p\text{-}n$ junctions in graphene using electronic transport, scanning tunneling microscopy, and first principles calculations. The $p\text{-}n$ junction lies at the boundary between differentially-doped regions of a graphene sheet, where one side is intrinsic and the other is charge-doped by proximity to a flake of \arucl\ across a thin insulating barrier. We extract the $p\text{-}n$ junction contribution to the device resistance to place bounds on the junction width. We achieve an ultra-sharp junction when the boundary between the intrinsic and doped regions is defined by a cleaved crystalline edge of \arucl\  located 2 nm from the graphene. Scanning tunneling spectroscopy in heterostructures of graphene, hexagonal boron nitride, and \arucl\ shows potential variations on a sub-10 nm length scale. First principles calculations reveal the charge-doping of graphene decays sharply over just nanometers from the edge of the \arucl\ flake.
\end{abstract}

Ideal $p\text{-}n$ junctions in graphene with a step-function change in carrier density underlie the physics of Klein tunneling \cite{katsnelson_chiral_2006,stander_evidence_2009,young_quantum_2009,allain_klein_2011}, negative refraction required for Veselago lensing \cite{cheianov_focusing_2007,chen_electron_2016}, guiding of plasmons \cite{mishchenko_guided_2010} and snake states \cite{williams_snake_2011}. Such junctions may also enable controlled anisotropy of the band velocity \cite{park_anisotropic_2008,li_anisotropic_2021}, novel electron-optical devices based on transformation optics \cite{vakil_transformation_2011}, or the ability to focus electron beams \cite{park_electron_2008,sajjad_high_2011,peterfalvi_intraband_2012,jang_graphene_2013}. In practice, $p\text{-}n$ junctions defined by electrostatic gating are far from this ideal, with the change in carrier density taking place over ${\approx}\,40\text{-}100\text{ nm}$ due to fringe electric fields from the edges of the metallic gates \cite{lee_observation_2015,chen_electron_2016,zhou_atomic-scale_2019}, precluding observation of effects such as Veselago lensing that rely on sub-10-nm junctions\cite{chen_electron_2016}. Here we employ modulation-doping of graphene by close (${\lesssim}2\text{ nm}$) proximity to \arucl\ \cite{wang_modulation_2020}, and a well-defined boundary to the doped region via a cleaved edge of the \arucl\ flake, to create ultra-sharp junctions, demonstrated with evidence from electronic transport, scanning tunneling probes, and first principles calculations.

When the layered Mott insulator alpha-ruthenium(III) chloride (\arucl) is placed in direct contact with graphene, it accepts approximately $4{\times}10^{13}\text{ cm}^{-2}$ electrons, leaving the graphene strongly hole-doped \cite{zhou_evidence_2019,mashhadi_spin-split_2019,gerber_ab_2020,biswas_electronic_2019,rizzo_charge-transfer_2020}. If an insulating spacer is introduced between the two materials, the charge transfer is weakened and the mobility increases commensurate with the setback of \arucl\ from graphene, analogous to modulation doping of conventional two-dimensional electron gases\cite{wang_modulation_2020,dingle_electron_1978}. The spatial distribution of the hole-doping is determined by where the \arucl\ overlaps the graphene which can in principle have an atomically-abrupt boundary. Thus charge-doping by \arucl\ appears to be a viable route toward ultra-sharp $p\text{-}n$ junctions in graphene.

Here we fabricate $p$-$n$ junctions in graphene by a combination of modulation-doping to differentially charge-dope two regions, and electrostatic gating to independently tune the densities in each. We use the resistance measured across the junctions to extract the junction width and find an ultra-sharp, ${\lesssim}10\text{ nm}$ junction when a cleaved crystalline edge of the dopant \arucl\ flake placed 2 nm away from the graphene defines the boundary between the regions. We further use low-temperature scanning tunneling microscopy and spectroscopy (STM/STS) to explore devices where a graphene sheet is either directly in contact with \arucl\ or separated from it by thin flakes of hexagonal boron nitride (hBN). We observe a sharp change in the charge doping of the graphene over a sub-$10\text{ nm}$ length scale across step edges in the insulating hBN spacer. Finally, we perform density functional theory (DFT) calculations that reveal the hole-doping of graphene due to electron transfer to \arucl\ falls off rapidly over just a few graphene lattice constants from the \arucl\ edge.

\begin{figure*}[t]
\includegraphics[width=\textwidth]{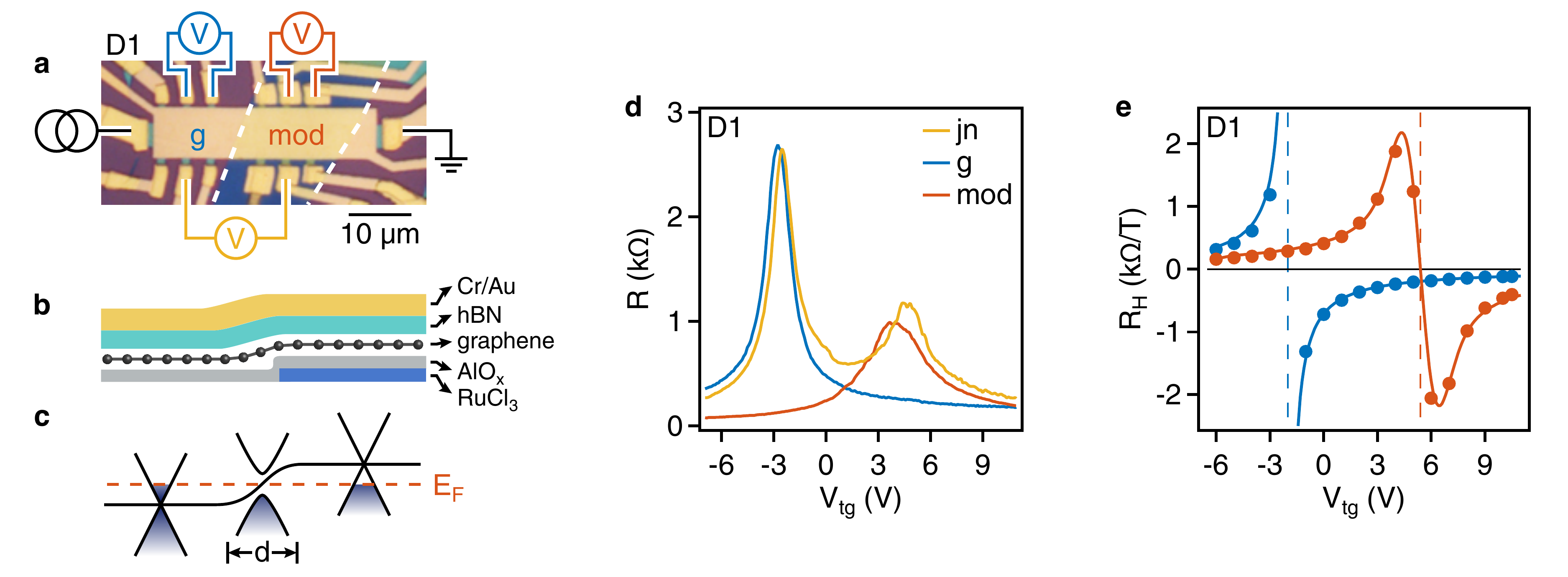}
\caption{\textbf{Spatial control of modulation doping in graphene.} \textbf{a} Optical micrograph of device. The white dashed lines indicate the \arucl\ flake boundary separating regions of intrinsic and \arucl-modulation-doped graphene, labeled ``g'' and ``mod'', respectively. \textbf{b} Schematic of device layer profile. \textbf{c} Schematic of the graphene band structure crossing from $n$-type to $p$-type across a junction of width $d$, showing an effective gap opening in the junction at the Fermi energy $E_F$. \textbf{d} Four-terminal resistance measured simultaneously in g (blue) and mod (orange) regions, and also across their interface (yellow), color-coded to the voltage measurement schematics of part \textbf{a}. \textbf{e} Low-field Hall coefficient of g and mod regions. Solid lines shows $R_{H} = -1/n e$ for the intrinsic (blue) and \arucl-doped (orange) regions; the latter is convolved with a Gaussian of width $\sigma=3.5\times10^{11}\text{ cm}^{-2}$ representing a spread of densities from electron-hole puddling.\cite{chandni_transport_2015} \label{ProxRuCl012_DP_Hall}} 
\end{figure*}

We present electronic transport in two graphene devices containing lateral $p\text{-}n$ junctions. In both, half the graphene sheet is intrinsic while the other half is modulation-doped by an \arucl\ flake. Device D1 has a ${\approx}1.5$-nm-thick AlO$_x$ film between the graphene and \arucl, while device D2 has a 2-nm-thick flake of hBN as a spacer. Figure \ref{ProxRuCl012_DP_Hall}\textbf{a} shows an optical microscope image of D1, which consists of a 16.5-nm-thick flake of \arucl\ coated by the AlO$_x$ film, topped by a graphene Hall bar that lies partly above the \arucl\ and partly on the bare substrate. The Hall bar is capped by a flake of hBN (${\approx}30\text{ nm}$) supporting a global Cr/Au top gate, and is contacted by Cr/Au leads. The entire device rests on $300\text{ nm}$ of SiO$_2$ on $p$-Si; the latter also serves as a global back gate. Further fabrication details are given in the Supporting Information \cite{supp}. In Fig.\ \ref{ProxRuCl012_DP_Hall}\textbf{a}, the device regions labeled ``g'' and ``mod'' correspond to the intrinsic and \arucl-doped graphene, respectively, and a profile of the device stack is shown schematically in Fig.\ \ref{ProxRuCl012_DP_Hall}\textbf{b}.

Four-terminal resistance measurements at $T{=}4$ K of the g and mod side of device D1 are shown in Fig.\ \ref{ProxRuCl012_DP_Hall}\textbf{d}, as a function of the top gate voltage. On either side of the junction we see resistance maxima at the graphene charge neutrality point (CNP or Dirac point), shifted by a few volts relative to each other due to the $p$-type modulation-doping \cite{zhou_evidence_2019,mashhadi_spin-split_2019,wang_graphene_2019}. Measurements of the low-field Hall coefficient, $R_H$, shown in Fig.\ \ref{ProxRuCl012_DP_Hall}\textbf{e} directly show the charge-doping difference to be $3.2{\times}10^{12}\text{ cm}^{-2}$. Similarly, in the hBN-spaced device D2, the charge transfer is $1.5\times10^{12}\text{ cm}^{-2}$ \cite{supp}. Typical g-side (mod-side) mobilities and mean free paths in D1 range from $8\text{,}000\text{-}12\text{,}000\text{ cm}^2/\text{Vs}$ and $100\text{-}250\text{ nm}$ ($6\text{,}000\text{-}10,000\text{ cm}^2/\text{Vs}$ and $50\text{-}200\text{ nm}$) \cite{supp}. Intriguingly, although the top and back gates are global, the carrier densities on either side can be independently tuned. Thus the back-gate electric field on the modulation-doped side must be screened by the \arucl\cite{supp}.

The width of a graphene $p\text{-}n$ junction can be determined by its contribution to the total device resistance. Charge carriers incident on a $p\text{-}n$ junction in graphene obey an electronic analog of Snell's law at an interface of right- and left-handed optical materials:\ the momentum along the junction, $k_y=k_F\sin\theta$, is conserved, but the momentum $k_x$ normal to the junction changes sign, the end result being a negative refraction \cite{cheianov_selective_2006,low_conductance_2009}. Here $k_F$ is the Fermi momentum and $\theta$ is the carrier angle of incidence on the junction. Carriers are transmitted across an abrupt junction with probability $T(\theta)=\cos^2\theta$ due to pseudospin conservation. In real devices there is always a density gradient from $p$- to $n$-type over some characteristic width $d$, analogous to the depletion region of a classical doped-Si $p\text{-}n$ junction. Although there is no band gap in graphene, an effective gap to transmission arises when $k_x(x)=\sqrt{(E(x)/\hbar v_F)^2-k_y^2}$ becomes imaginary, where $E(x)=\hbar v_F k_F$ is the position-dependent energy of the graphene Dirac point across the junction, and $v_F {\approx} 10^6$ m/s is the Fermi velocity. This is depicted schematically in Fig.\ \ref{ProxRuCl012_DP_Hall}\textbf{c}. Tunneling across this gap reduces the transmission probability as a function of impact angle and junction width which, for a balanced junction ($|p|{=}|n|$), is given by \cite{cheianov_selective_2006}

\begin{equation}
  T(\theta,d) = \cos^2\theta e^{-\pi k_F d \sin^2\theta}~. \label{tbal}
\end{equation}

\noindent The reduced transmission leads to a finite resistance that has both ballistic and diffusive contributions, $R_{p\text{-}n}{=}R_{bal} + R_{dif}$, whose relative magnitudes depend on the carrier mean free path and also many-body effects \cite{zhang_nonlinear_2008,fogler_effect_2008,supp}. Experimental values of $R_{p\text{-}n}$ range from a few hundred ohms in graphene-on-oxide junctions to 100 $\Omega$ in hBN-encapsulated junctions \cite{huard_transport_2007,stander_evidence_2009,chen_electron_2016}. 

\begin{figure*}[t!]
\includegraphics[width=\textwidth]{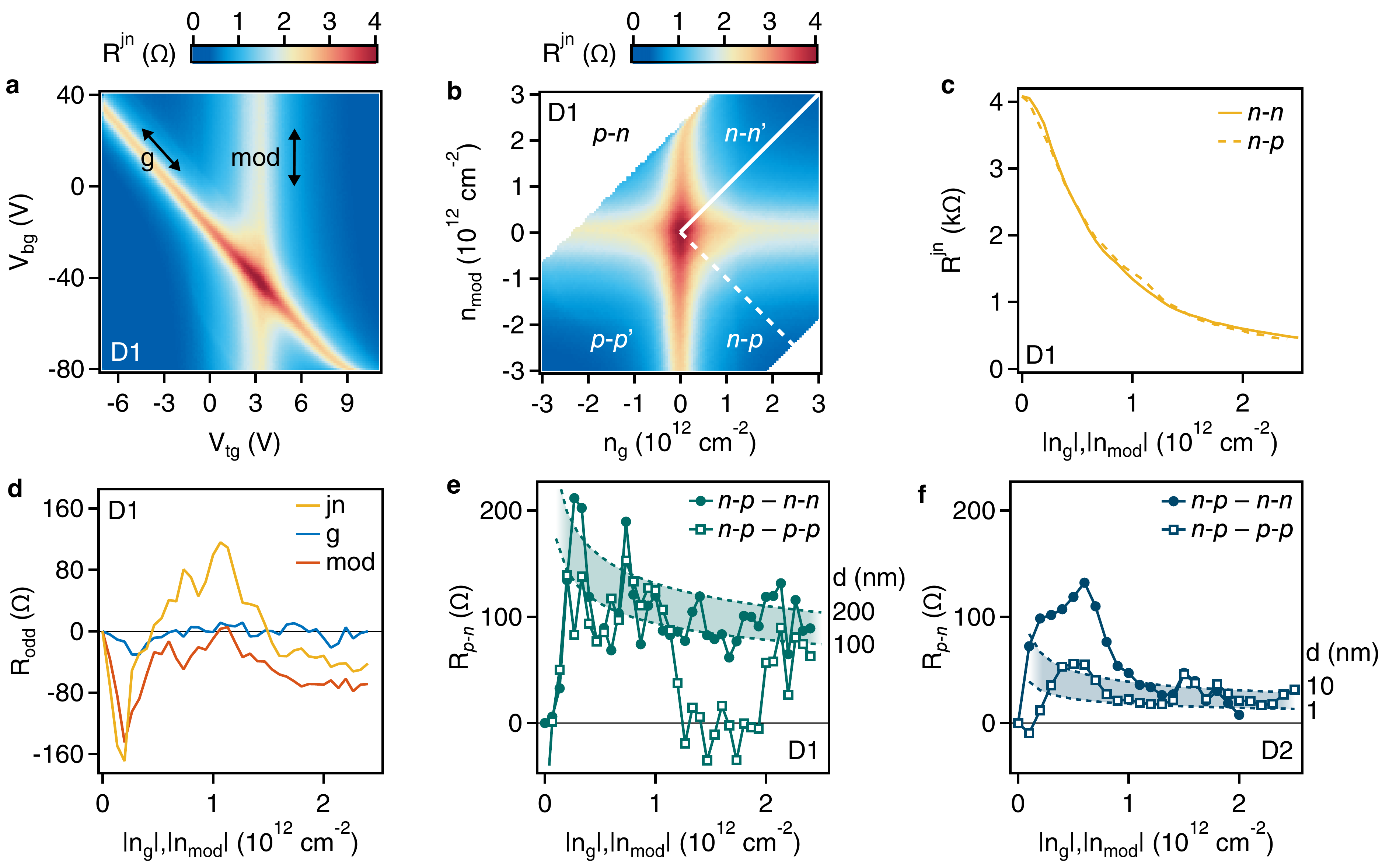}
\caption{\textbf{Resistance across a modulation-doping-defined $p\text{-}n$ junction.} \textbf{a} Four-terminal resistance across the interface of intrinsic (g) and modulation-doped (mod) graphene, as a function of top and back gate voltages. \textbf{b} Same as \textbf{a}, now re-plotted as a function of the g- and mod-side carrier densities. The labels show the polarity of the four quadrants defined by the charge neutrality peaks, either monopolar (e.g.\ $n\text{-}n'$) or bipolar ($n\text{-}p$). White solid and dashed lines mark where the carrier density on either side of the interface is equal ($n$-$n$), or of equal magnitude but opposite sign ($p\text{-}n$). \textbf{c} Comparison of resistances at the white solid and dashed lines in \textbf{b}. \textbf{d} Difference of $R^{jn}$ (yellow) for the two curves in \textbf{b}; and of the g-side resistance $R^g$ (blue) and mod-side resistance $R^{mod}$ (orange) for line cuts at the same carrier densities (or gate voltages). \textbf{e} $p\text{-}n$ junction resistance for device D1. Shaded region marks the theoretical resistance for a ballistic device with junction width ranging between 100 and 200 nm. \textbf{f} $p\text{-}n$ junction resistance for device D2. Shaded region marks the theoretical resistance for a ballistic device with junction width ranging between 1 and 10 nm. \label{ProxRuCl012_Rpn}}
\end{figure*}

We extract the width of lateral $p\text{-}n$ junctions in two devices as follows, illustrating the procedure by analyzing the transport in device D1. First, in Fig.\ \ref{ProxRuCl012_Rpn}\textbf{a} we show the top- and back-gate dependence of the total resistance across the junction, $R^{jn}$, using the contacts marked in yellow in Fig.\ \ref{ProxRuCl012_DP_Hall}\textbf{a} for D1. This quantity includes the sheet resistances from both sides of the junction, and $R_{p\text{-}n}$ due to the junction itself. The density of the intrinsic portion of the graphene depends on both gates as $n_{g} = \alpha_{tg} V_{tg} - \alpha_{bg} V_{bg} + n_{g,0}$, while the \arucl-doped portion depends only on the top gate via $n_{mod} = \alpha_{tg} V_{tg}+n_{mod,0}$. Here the top and back gating efficiencies are $\alpha_{tg}=4.4\times10^{11}\text{ cm}^{-2}$/V and $\alpha_{bg}=6.0\times10^{10}\text{ cm}^{-2}$/V, respectively, and the densities for zero applied gate bias are $n_{g,0}=1.1\times10^{12}\text{ cm}^{-2}$ and $n_{mod,0}=-1.3\times10^{12}\text{ cm}^{-2}$. Using these relations, in Fig.\ \ref{ProxRuCl012_Rpn}\textbf{b} we re-plot $R^{jn}$ vs the carrier densities of the intrinsic graphene, $n_{g}$, and the modulation-doped graphene, $n_{mod}$. The CNPs of these two differentially-doped regions appear as vertical and horizontal bands separating the regimes of same-sign ($p$-$p'$ or $n$-$n'$) carrier transport across the interface from those having bipolar transport.

Next we isolate the resistance, $R_{p\text{-}n}$, of the $p\text{-}n$ junction itself. We start with line cuts of $R^{jn}$ along lines of equal carrier density and same sign ($n_g{=}n_{mod}>0$, yielding $R^{jn}_{n\text{-}n}$) or opposite sign ($n_g{=}-n_{mod}>0$, $R^{jn}_{n\text{-}p}$). These are plotted together in Fig.\ \ref{ProxRuCl012_Rpn}\textbf{c}. To the extent that sheet resistances on either side of the interface are symmetric with respect to charge neutrality, then the difference of these curves, $R_{odd}^{jn}{=}R^{jn}_{n\text{-}p}-R^{jn}_{n\text{-}n}$, will be due \textit{only} to the resistance of the $p\text{-}n$ junction:\ $R_{odd}^{jn}{=}R_{p\text{-}n}$. In fact, for the line cuts in Fig.\ \ref{ProxRuCl012_Rpn}\textbf{c} the contribution from the g-side sheet resistance ought to be identical because the g-side carrier density does not change sign. In contrast, the two line cuts include either $n$- or $p$-type doping of the mod side, so any asymmetry about the CNP in this region will add an additional resistance to $R_{odd}^{jn}$ that must be subtracted off. To determine the presence of this additional contribution, we make resistance maps analogous to Fig.\ \ref{ProxRuCl012_Rpn}\textbf{b} for both the g and mod side \cite{supp}. From equivalent line cuts along the $n\text{-}n$ and $n\text{-}p$ directions, we calculate $R_{odd}^g$ and $R_{odd}^{mod}$. These, along with $R_{odd}^{jn}$, are shown in Fig. \ref{ProxRuCl012_Rpn}\textbf{d}. As expected, $R_{odd}^g$ lies close to zero; but $R_{odd}^{jn}$ and $R_{odd}^{mod}$ are finite and share a similar lineshape. 

It remains to subtract this asymmetric part of the sheet resistance to finally obtain the $p\text{-}n$ junction resistance:\ $R_{p\text{-}n}{=}R_{odd}^{jn}-(c_g{\times}R_{odd}^g+c_{mod}{\times}R_{odd}^{mod})$, where $c_i$ are scaling factors appropriate to the device geometry \cite{supp}. Figure \ref{ProxRuCl012_Rpn}\textbf{e} shows the resulting $R_{p\text{-}n}$ values, along with the analysis for line cuts along the $(n\text{-}p)$ and $(p\text{-}p)$ directions which should in principle yield the same junction resistance. Indeed, both show values of ${\sim}100\ \Omega$ (but for a brief excursion by the $(p\text{-}p)$-derived trace which can be attributed to a dip in $R^{jn}$\cite{supp}). Figure \ref{ProxRuCl012_Rpn}\textbf{f} shows the results of similar analyses carried out in device D2. Here, both curves show peaks at low density that rapidly converge to values between 20 and 30 $\Omega$ over much of the carrier density range. We compare these results to theoretical predictions for the resistance of $p\text{-}n$ junctions in disordered graphene \cite{fogler_effect_2008}, which we plot as shaded bands calculated for junction widths $d$ that span 100 to 200 nm and 1 to 10 nm in Fig.\ \ref{ProxRuCl012_Rpn}\textbf{e} and \textbf{f},  respectively. Device D2 is thus found to have an ultra-sharp, sub-10-nm junction, while D1 has a much wider ${\sim}100\text{ nm}$ junction.

At first, this result is surprising:\ why are the two junction widths so different? Both have insulating spacers of approximately the same thickness, with modulation-doping levels only a factor of two apart. The differing mobilities are unlikely to be the culprit, as transport across the junction is firmly in the ballistic regime\cite{supp}. The interface in device D1 is angled at $22^{\circ}$ so the junction appears wider, but only by a factor of $1/\cos(22^{\circ})\approx 1.08$ \cite{sajjad_manipulating_2013}. Ultimately, inspection of the \arucl\ flakes used in the devices offers a clear resolution:\ in D1, the edge of the \arucl\ flake at the boundary between the intrinsic and modulation-doped regions is slightly curved, with no obvious relation to its crystalline axes. In contrast, for D2 the edge defining the boundary of charge transfer is straight and makes an angle of ${\approx}119^{\circ}$ with another portion of the flake just outside where it contacts the graphene\cite{supp}. This implies the boundary in D2 is a cleaved crystalline edge, and in D1 is likely to be rough with various facets along the edge.

\begin{figure*}[t!]
\includegraphics[width=\textwidth]{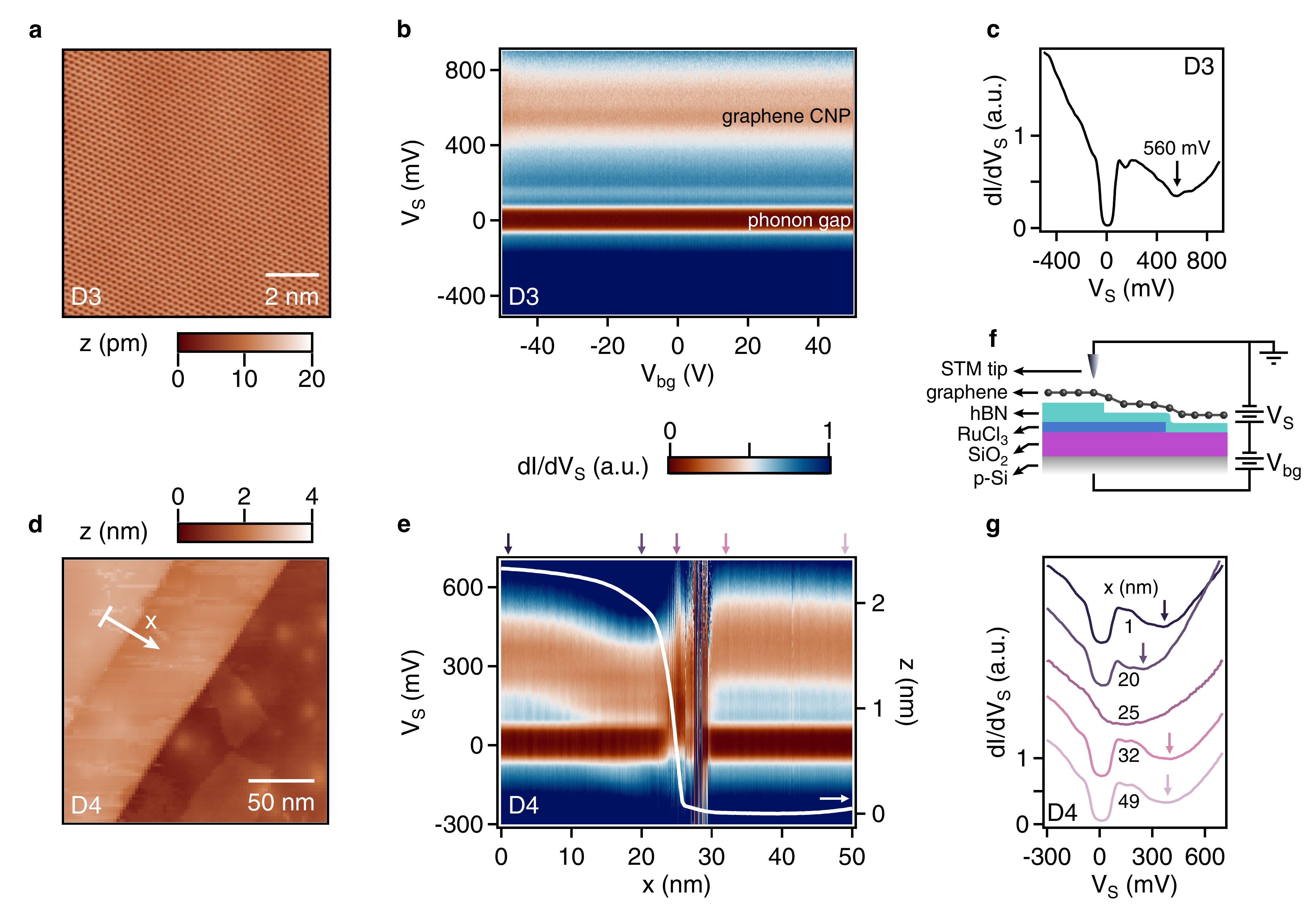}
\caption{\textbf{STM and STS across step edges in graphene/hBN/\arucl\ heterostructures}. \textbf{a} Atomically resolved STM topography of a $10{\times}10\text{ nm}^2$ region taken on graphene/\arucl\ in Device D3 at $T{=}4.8$ K. \textbf{b} Color map of the measured d$I$/d$V_S$ ($V_S$,d) from region in \textbf{a}, as a function of gate voltage applied to substrate. The strong feature near $V_S{=}0$ mV corresponds to phonon-assisted inelastic tunneling while the additional suppression in intensity near $V_S{=}560$ mV corresponds to the graphene charge neutrality point. \textbf{c} Average of spectra across all $V_{bg}$ shows a clear phonon gap and minimum at CNP. \textbf{d} STM topography of a $200\times200\text{ nm}^2$ window in a graphene/hBN/\arucl\ region in Device D4. \textbf{e} Color map of the measured d$I$/d$V_S$ along the white line over a step edge in \textbf{d}. The phonon gap and graphene CNP are readily visible, with the latter showing a non-monotonic dispersion as the tip travels over the step edge. White line tracks the change in tip height in crossing the step. \textbf{f} Schematic of measurement over step edges in graphene/hBN/\arucl/SiO$_2$/Si heterostructure. \textbf{g} Measured d$I$/d$V_S$ spectra at various $x$ positions (labeled in figure) along white line in \textbf{d}, highlighting the non-monotonic shift of the CNP feature, indicated by an arrow for each curve.} 
\label{stm}
\end{figure*}

We use scanning tunneling microscopy and spectroscopy at $T{=}4.8$ K to study the spatial variation of the Dirac point across differentially-doped regions in two other devices, D3 and D4, both composed of overlapping flakes of graphene, hBN, and \arucl\ on a SiO$_2$/$p$-Si substrate. Figure \ref{stm}\textbf{a} shows an atomically resolved topographic map of a region in D3, consisting of graphene in direct contact with \arucl. The differential tunneling current, d$I$/d$V_S$, proportional to the local density of states (LDOS), is acquired as a function of the tip-sample bias, $V_S$, and a back gate voltage applied to the substrate, $V_{bg}$, with results plotted in Fig.\ \ref{stm}\textbf{b} where a dark blue (brown) color corresponds to high (low) LDOS. A strong dark brown band centered about $V_S{=}0$ mV appears along with several fainter features. We show the averaged spectra from $V_{bg} {=} {-}50$ V to ${+}50$ V in Fig.\ \ref{stm}\textbf{c} which shows a 120-mV-wide U-shaped suppression of d$I$/d$V_S$ centered about $V_S {=}0$ mV, with a less pronounced minimum at $V_S {=} 560$ mV. The former is a familiar phonon-assisted inelastic tunneling gap \cite{zhang_giant_2008}, while the latter corresponds to the graphene CNP \cite{zhang_giant_2008,zhang_origin_2009}. We estimate the graphene carrier density using $n_g=(E_{DP}-\hbar \omega)^2/(\pi \hbar^2 v_F^2)$, with $\hbar \omega$ the phonon energy and $E_{DP}$ the energy of the Dirac point in Fig.\ \ref{stm}\textbf{c}, and find a large $p$-type doping of $n{=}{-}1.8{\times}10^{13}\text{ cm}^{-2}$, on the low side of prior observations of the graphene/\arucl\ charge transfer\cite{zhou_evidence_2019,wang_modulation_2020,rizzo_charge-transfer_2020}. The surprising lack of response to the back gate corroborates the screening effect noted above in transport for \arucl-doped graphene.

In Fig.\ \ref{stm}\textbf{d}, we show a topographic map of a region in device D4 that shows terraces due to separation of graphene and \arucl\ by an hBN spacer of varying thickness, shown schematically in Fig.\ \ref{stm}\textbf{f}. In Fig.\ \ref{stm}\textbf{e}, we show d$I$/d$V_S$ spectra acquired over one such edge (along the arrow in Fig.\ \ref{stm}\textbf{d}) as a function of both $V_S$ and position $x$ to map the change in charge transfer. The white curve shows the height profile (right axis). As above, the phonon gap appears at $V_S{=} 0$ mV, but the graphene CNP feature disperses non-monotonically with $x$, briefly decreasing as the step edge is approached and then sharply increasing to a final plateau once the step edge is crossed. Point spectroscopy taken at different $x$ values (shown in Fig.\ \ref{stm}\textbf{g}) illustrates the non-monotonic variation of the CNP across the step edge. Far from the edge, the charge density is found to be $5.8{\times}10^{12}\text{ cm}^{-2}$ ($8.3{\times}10^{12}\text{ cm}^{-2}$) for the higher (lower) step, confirming that larger charge transfers are associated with thinner hBN spacer layers. The positive shift of the CNP takes place rapidly over ${\approx}7\text{ nm}$. A recent work in which STM is used to map the charge density in a nanobubble in graphene on \arucl, at room temperature, finds an even sharper interface across a $p\text{-}n$ junction\cite{rizzo_nanometer-scale_2022}. We note an instability in STS is observed at the step edge, where the tip-sample interaction may lead to a small delamination of the graphene with a decrease in the charge transfer \cite{klimov_electromechanical_2012}.

\begin{figure*}[t!]
\includegraphics[width=\textwidth]{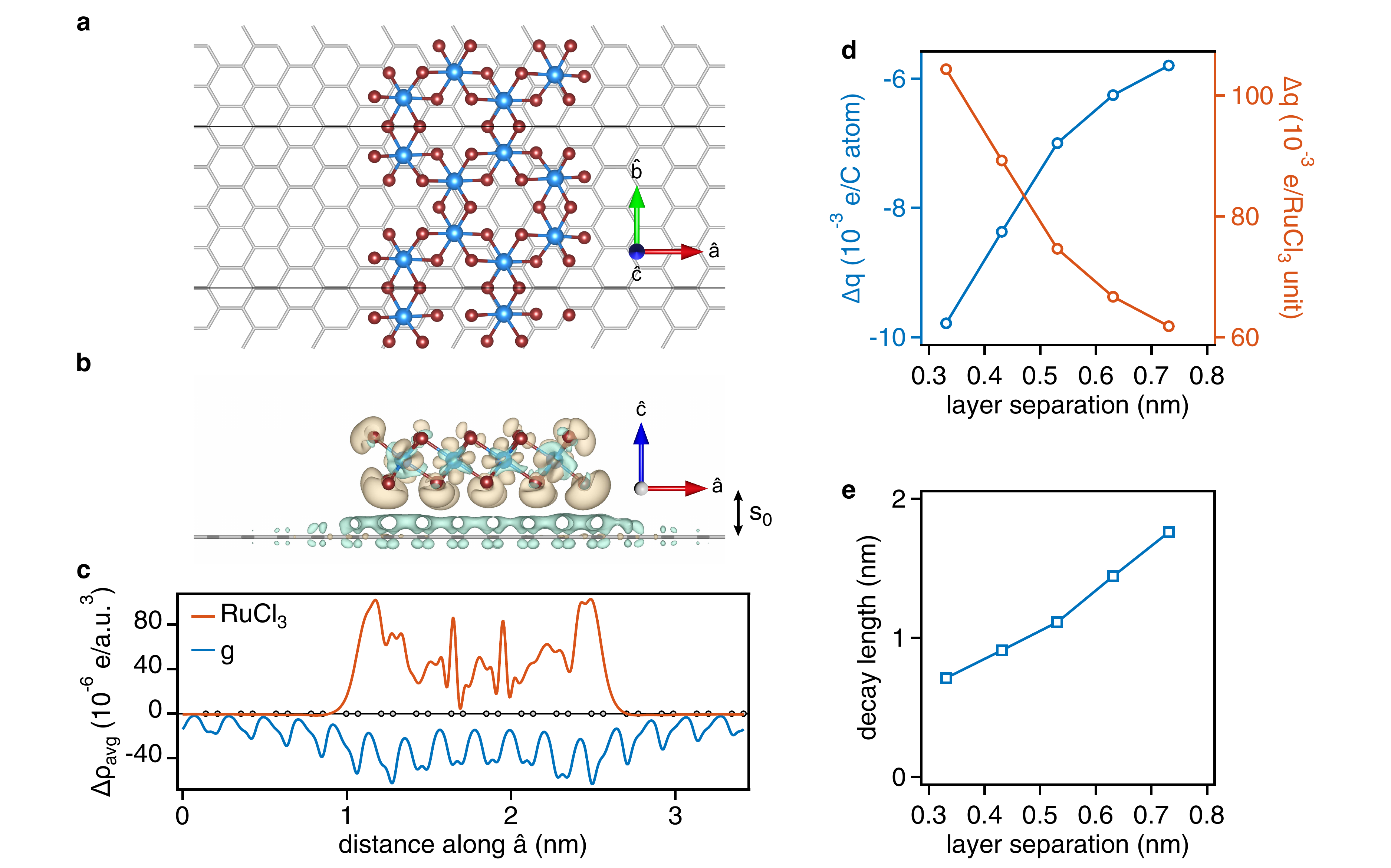}
\caption{\textbf{Calculated charge densities of \arucl\ ribbon on graphene.} \textbf{a} Top view of the supercell and the corresponding orthogonal directions of the lattice vectors ($\hat{a}, \hat{b}$, and $\hat{c}$). The blue (red) spheres represent the Ru (Cl) atoms. \textbf{b} Side view of supercell with illustration of the change in charge density, $\Delta \rho$, showing accumulation and depletion in beige and teal, respectively. The isosurface value of $\Delta \rho$ is chosen to be $5 {\times} 10^{-4}\text{ \textit{e}/a.u.}^3$, and $s_0=3.31\text{ \AA}$ is the equilibrium separation of the relaxed geometry. \textbf{c} Planar average of the change in charge density (in the $\hat{b}$, $\hat{c}$ plane) along the $\hat{a}$ direction, $\Delta \rho_{avg}$, separated by color for charge on \arucl\ (orange) and graphene (blue). \textbf{d} Total integrated charge on \arucl\ (orange, right axis) and graphene (blue, left axis) as a function of separation between the layers. \textbf{e} Decay length of charge distribution in graphene as a function of lateral distance away from the edge of the \arucl\ ribbon.} 
\label{theo}
\end{figure*}

To understand both the lateral and vertical spatial distribution of the charge transfer due to the modulation-doping of graphene by \arucl, we perform first principles calculations of a monolayer-thick \arucl\ ribbon on graphene as shown in Fig.~\ref{theo}\textbf{a} and \textbf{b}. By using DFT calculations as implemented in VASP\cite{supp}, we first calculate the properties of the interface when no spacer layers are present in a large supercell in the ribbon-on-sheet geometry (see Fig.~\ref{theo}\textbf{a}), with supercell lattice parameters $a$=34.16 \AA, $b$=9.84 \AA\ and $c$=20 \AA\  (distance between periodic images along $c$ direction is $\approx$ 16.5 \AA), and the graphene C-C bond-lengths fixed at 1.42 \AA. Geometrical optimization of the internal atomic degrees of freedom leads to a mildly distorted Ru hexagon with shorter Ru-Ru bonds  ($l_s$=3.17 \AA) arranged in a periodic-step-function like pattern along the direction $\hat{b}$, compared with the other Ru-Ru bonds of the ribbon ($l_s$=3.48-3.52 \AA).

With the two materials in close proximity, a new charge density distribution develops which we illustrate by subtracting of the charge densities of intrinsic graphene and \arucl\ from that of the graphene/\arucl\ heterostructure, $\Delta \rho = \rho_{\alpha\mathrm{R}/\mathrm{g}} - \rho_{\alpha\mathrm{R}} - \rho_{\mathrm{g}}$. We find that charge accumulates in the \arucl\ ribbon with a concomitant depletion in the graphene, as shown in Fig.~\ref{theo}\textbf{b} where we plot the charge isosurface at $|\Delta \rho|=5{\times}10^{-4}~ e$/a.u.$^{3}$, and in Fig.\ \ref{theo}\textbf{c} by directly plotting the variations in the planar average of $\Delta \rho$ (over $\hat{b}{\times}\hat{c}$) along the $\hat{a}$ direction for the graphene and \arucl\ layers. These results are in accord with findings for the graphene/\arucl\  commensurate bilayer case \cite{biswas_electronic_2019}. Figure \ref{theo}\textbf{c} shows the excess electronic charge in \arucl\ tends to lie largely on the Cl atoms facing the graphene. The majority of the charge depletion in graphene is concentrated at the C atom locations underneath the \arucl, reaches maxima near the boundaries of the \arucl\ ribbon, and proceeds to decrease rapidly beyond the edge. Adding a second \arucl\ layer does not qualitatively alter this result \cite{supp}. 

The equilibrium height of the \arucl\ above the graphene, $s_0{=}3.31$ \AA, is defined as the average distance between the C atoms in graphene and the graphene-facing Cl atoms in \arucl, shown in Fig.\ \ref{theo}\textbf{b}. To mimic the presence of a dielectric spacer layer, we calculate how the charge transfer changes if the separation is increased up to an additional 4 \AA\ (without further relaxing the geometry, but with the supercell lattice parameter $c$ also increased up to 24 \AA). The results, in Fig.\ \ref{theo}\textbf{d}, show a clear decrease of the charge exchange between the two layers, in qualitative agreement with the experimentally observed modulation doping effect. Performing the calculation with a dielectric present would alter the absolute magnitude of charge transfer but is expected to keep the relative changes similar to what we have found here \cite{wang_modulation_2020}.

Finally, we can estimate the characteristic length scale over which the charge transfer decays away outside the ribbon by fitting the decrease of the charge density peaks around the C atoms, visible in Fig.\ \ref{theo}\textbf{c}, as a function of distance. We find the best fit to the data is made using an equation of the form $A e^{-(x-x_0)/B}$, with $x_0$ measured along $\hat{a}$ from the average position of the zigzag C atoms just outside the \arucl\ ribbon. The decay length, $B$,  plotted in Fig.\ \ref{theo}\textbf{e} is an average of the fits made on either side of the \arucl\ ribbon, and is found to be roughly 2.5 times greater than the graphene/\arucl\ separation \cite{supp}.

We have demonstrated ultra-sharp $p\text{-}n$ junctions in modulation-doped graphene devices. This innovation relies upon several advantages conferred by using \arucl\ to charge dope graphene.

First, we use a cleaved crystalline edge to define an atomically-sharp and straight interface along the several-micron-length of the $p$-$n$ junction. Prior work has determined that roughness of this interface can be a significant hurdle to achieving ultra-sharp junctions\cite{elahi_impact_2019}. This advantage is not unique to \arucl; however, even in cleaved graphite-gate-defined $p$-$n$ devices with sub-nm lateral roughness along the interface, 40-nm-wide junctions still are observed\cite{zhou_atomic-scale_2019}. Thus a sharp interface may be necessary but is apparently not sufficient to obtain ultra-sharp junctions.

Therefore we note additional advantages unique to the \arucl\ approach. Our devices are composed of two monolayer charge distributions (only the layer of \arucl\ closest to graphene is appreciably charged \cite{wang_modulation_2020,supp}). These form a nearly ideal parallel-plate capacitor geometry with arbitrarily small separation between the plates. Indeed, the charge distribution in the \arucl\ shows accumulation near the edge of the ribbon, visible in Fig.\ \ref{theo}\textbf{c}, just as expected for a classical charged sheet over a metallic plane. This suggests the extent of the potential variation in graphene beyond the edge of \arucl\ (e.g.\ the junction width) is, apart from a possible role for nonlinear screening \cite{zhang_nonlinear_2008}, essentially a matter of electrostatics and thus due to the usual fringing electric fields which have a lateral extent on the order of the plate separation. Given this, it should be possible to achieve a similar result by implementing a graphite gate just as close to the graphene; however, this presents numerous practical difficulties including dielectric breakdown, unwanted leakage currents to the conducting gate, and the onset of tunneling for dielectric thicknesses below 2 nm\cite{britnell_electron_2012}. These are \textit{not} limitations for \arucl-doped graphene, for which the charge transfer is fixed, requires no external bias, and crucially has no leakage current due to the insulating nature of \arucl.

Thus the narrowest $p\text{-}n$ junctions can be achieved by placing a flake of \arucl\ with a cleaved edge as close as possible to the graphene. Junctions defined in this manner should be narrow enough to enable observation of electron-optical effects such as Veselago lensing and other useful devices based on electron refraction or reflection\cite{chen_electron_2016}.

\begin{suppinfo}

Additional information on sample preparation and device fabrication, device mobilities and mean free paths, identification of \arucl\ crystallographic edges, analysis of junction resistance data, scanning tunneling measurements, details of density functional theory calculations, and screening of back gate by \arucl.

\end{suppinfo}

\begin{acknowledgement}

We acknowledge enjoyable and informative discussions with K.\ Burch, D.\ Basov, Y.\ Wang, E.\ Gerber, and D.\ Rizzo. Support for device fabrication and measurement by the Institute of Materials Science and Engineering at Washington University in St.\ Louis is gratefully acknowledged. E.A.H., J.B., and J.B.\ received partial support from the National Science Foundation under DMR-1810305, and E.A.H.\ was additionally supported by NSF CAREER DMR-1945278. S.L.\ was supported by Office of Naval Research 6.1 Base Funding and the Jerome and Isabella Distinguished Scholar Fellowship. K.W.\ and T.T.\ acknowledge support from the Elemental Strategy Initiative
conducted by the MEXT, Japan (Grant Number JPMXP0112101001) and  JSPS KAKENHI (Grant Numbers 19H05790, 20H00354 and 21H05233). J.V.J.\ and Z.G.\ acknowledge support from the National Science Foundation under award DMR-1753367. J.V.J.\ acknowledges support from the Army Research Office under contract W911NF-17-1-0473. S.B.\ and R.V.\ thank the Deutsche Forschungsgemeinschaft (DFG, German Research Foundation)-TRR 288-422213477 (Project A05). M.C. and D.M. were supported by the National Science Foundation, Grant No.\ DMR-1808964.
\end{acknowledgement}

\bibliography{pn_jn}

\newpage


\section*{Supplementary material}

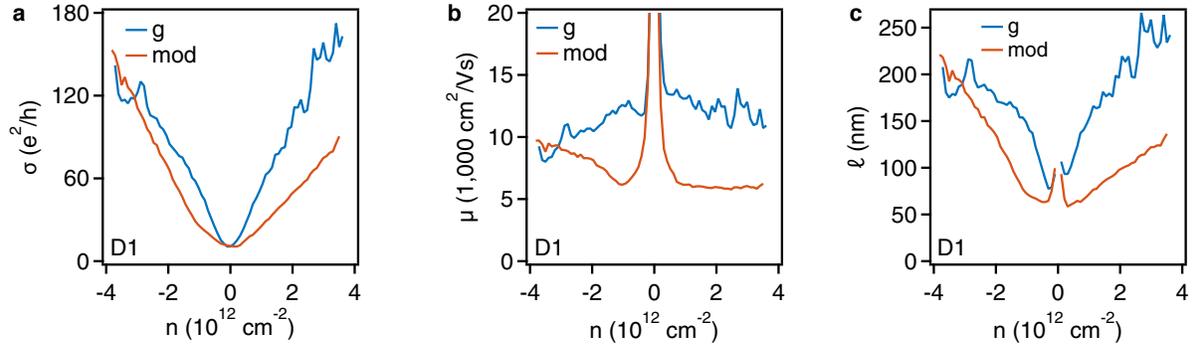

Figure S1: **Electronic transport properties of device D1 (AlO$_x$ spacer). a** Conductivities of the g and mod regions of device D1. **b** Mobilities of D1. **c** Mean free paths of D1.

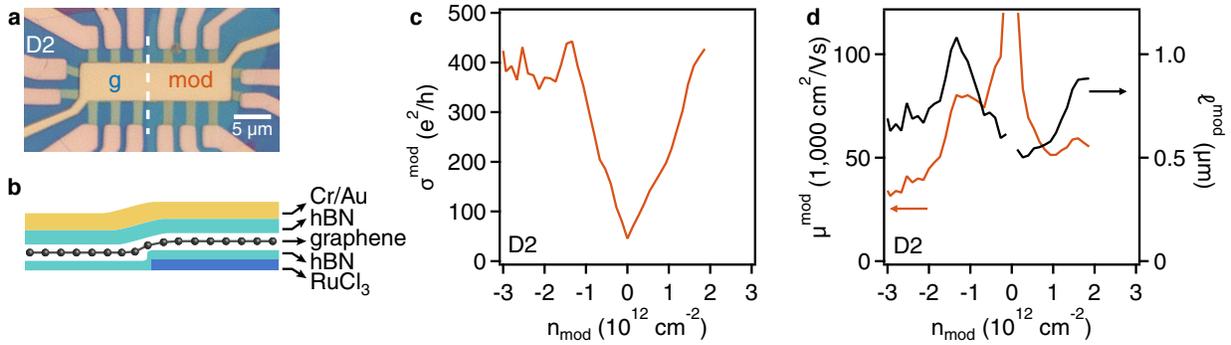

Figure S2: **Electronic transport properties of device D2 (hBN spacer). a** Optical micrograph of device D2. **b** Schematic of the stacking arrangement. **c** Conductivity of the mod side of D2. **d** Mobility (orange) and mean free path (black) of D2.

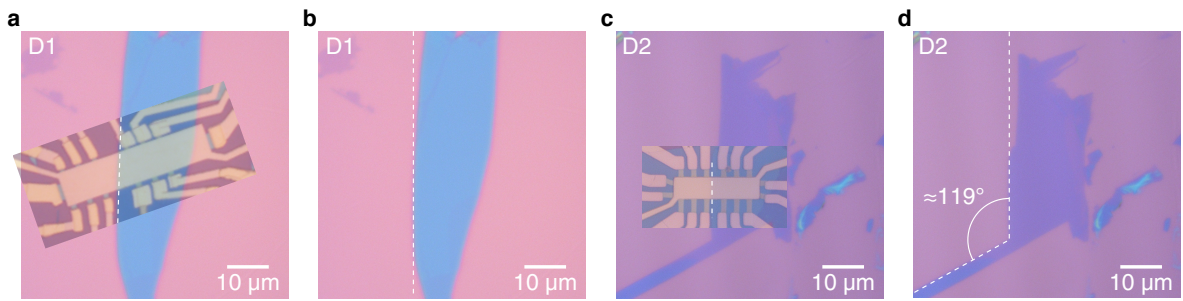

Figure S3: **$\alpha$-RuCl$_3$ flake edges. a** Optical microscope image of the $\alpha$-RuCl$_3$ flake used in device D1, with the final device image overlaid as a visual aid. **b** The $\alpha$-RuCl$_3$ flake from **a** with a vertical dashed line indicating the edge used to form the junction interface. **c** Optical microscope image of the $\alpha$-RuCl$_3$ flake used in device D2, with the final device image overlaid as a visual aid. **d** The $\alpha$-RuCl$_3$ flake from **c** with a vertical dashed line indicating the edge used to form the junction interface, measured at an angle of $\approx$119° from another flake edge.

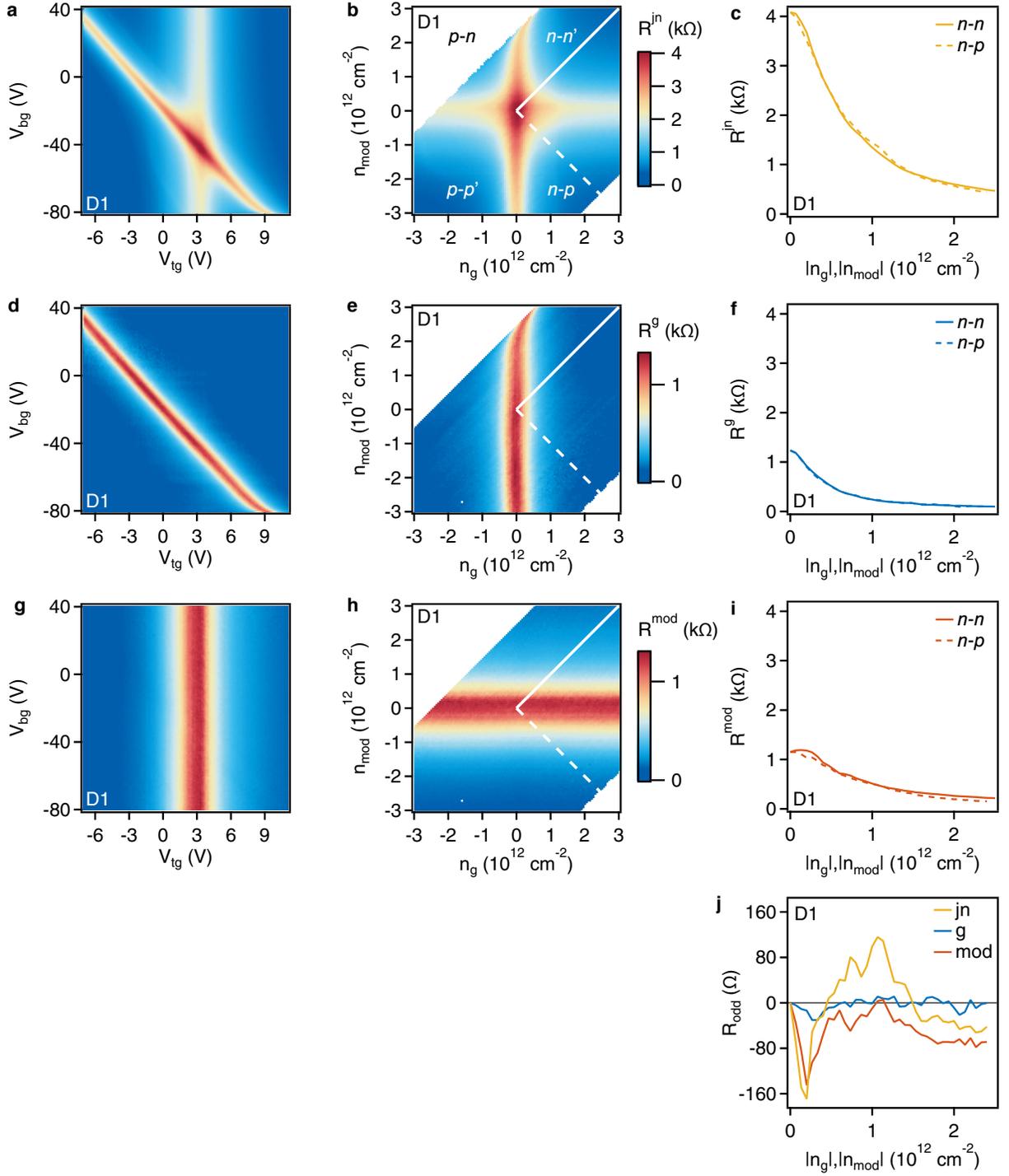

Figure S4: **Comparison of $n$-$n$ and $n$-$p$ resistances in device D1. a,d,g** Resistance across the junction, and in the g and mod sides as a function of top and bottom gates. **b,e,h** Resistances from **a,d,g** re-plotted as a function of g- and mod-side densities. **c,f,i** Linecuts along solid and dashed lines from **b,e,h**. **j** Differences between the curves in **c,f,i**. Panels **a**, **b**, and **j** are identical to Fig. 2**a**, **b**, and **d** of the main text.

S11

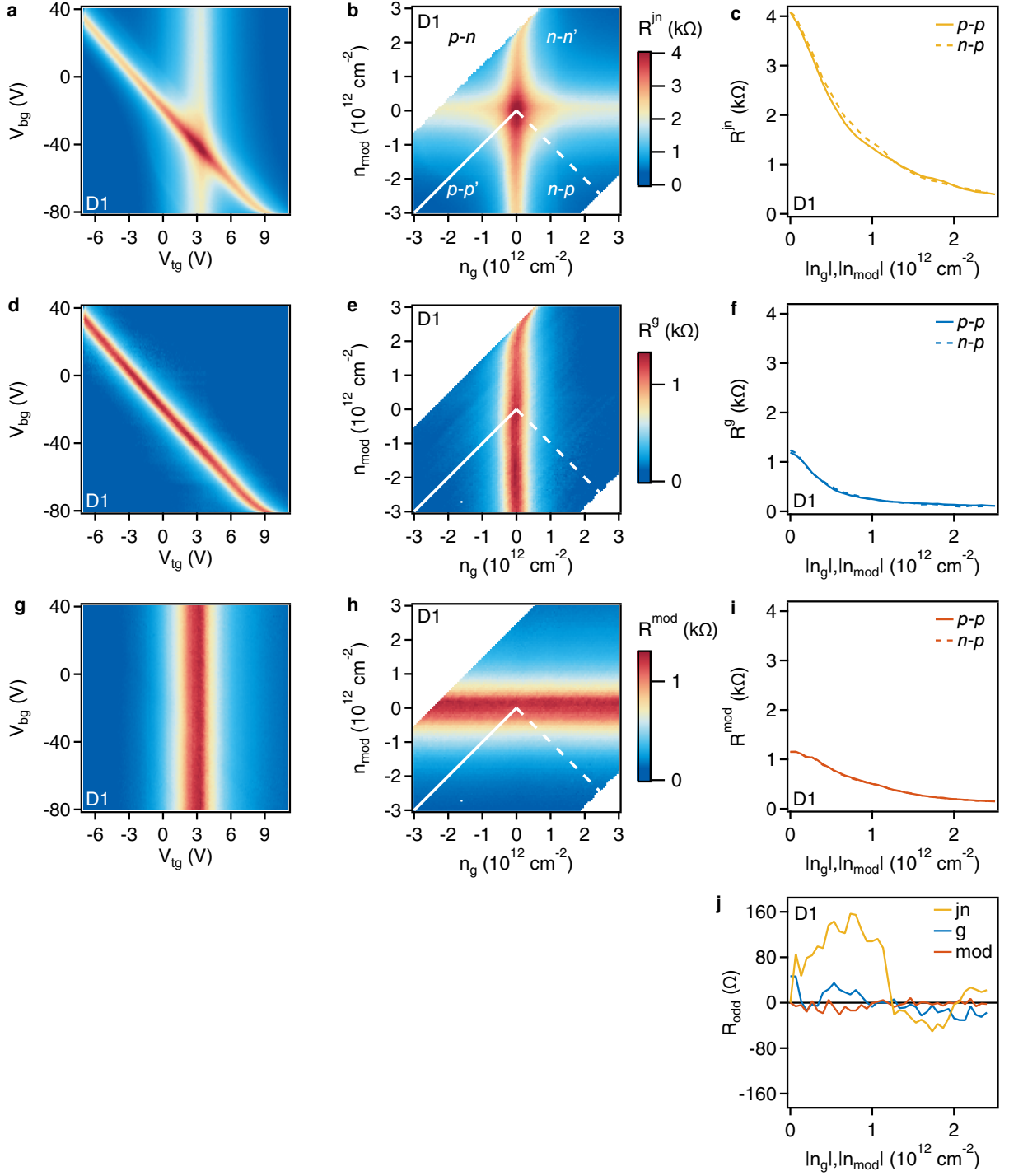

Figure S5: **Comparison of *p-p* and *n-p* resistances in device D1. a,d,g** Resistance across the junction, and in the g and mod sides as a function of top and bottom gates. **b,e,h** Resistances from **a,d,g** re-plotted as a function of g- and mod-side densities. **c,f,i** Linecuts along solid and dashed lines from **b,e,h**. **j** Differences between the curves in **c,f,i**. Panels **a** and **b** are identical to Fig. 2a and b of the main text (except for indicating different line cuts).

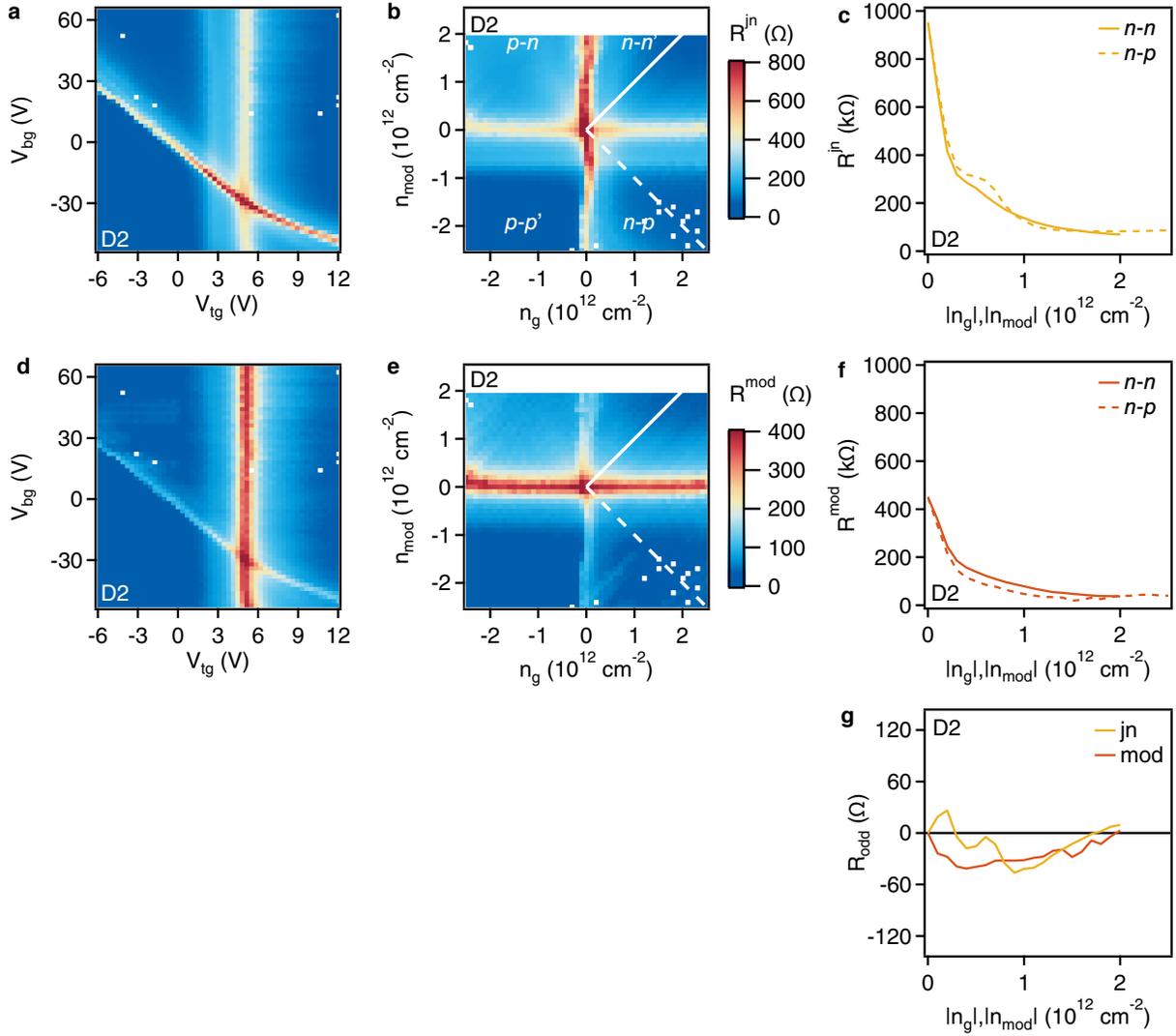

Figure S6: **Comparison of $n$-$n$ and $n$-$p$ resistances in device D2. a,d** Resistance across the junction and in the mod side as a function of top and bottom gates. **b,e,h** Resistances from **a,d,g** re-plotted as a function of g- and mod-side densities. **c,f** Linecuts along solid and dashed lines from **b,e**. **j** Differences between the curves in **c,f**.

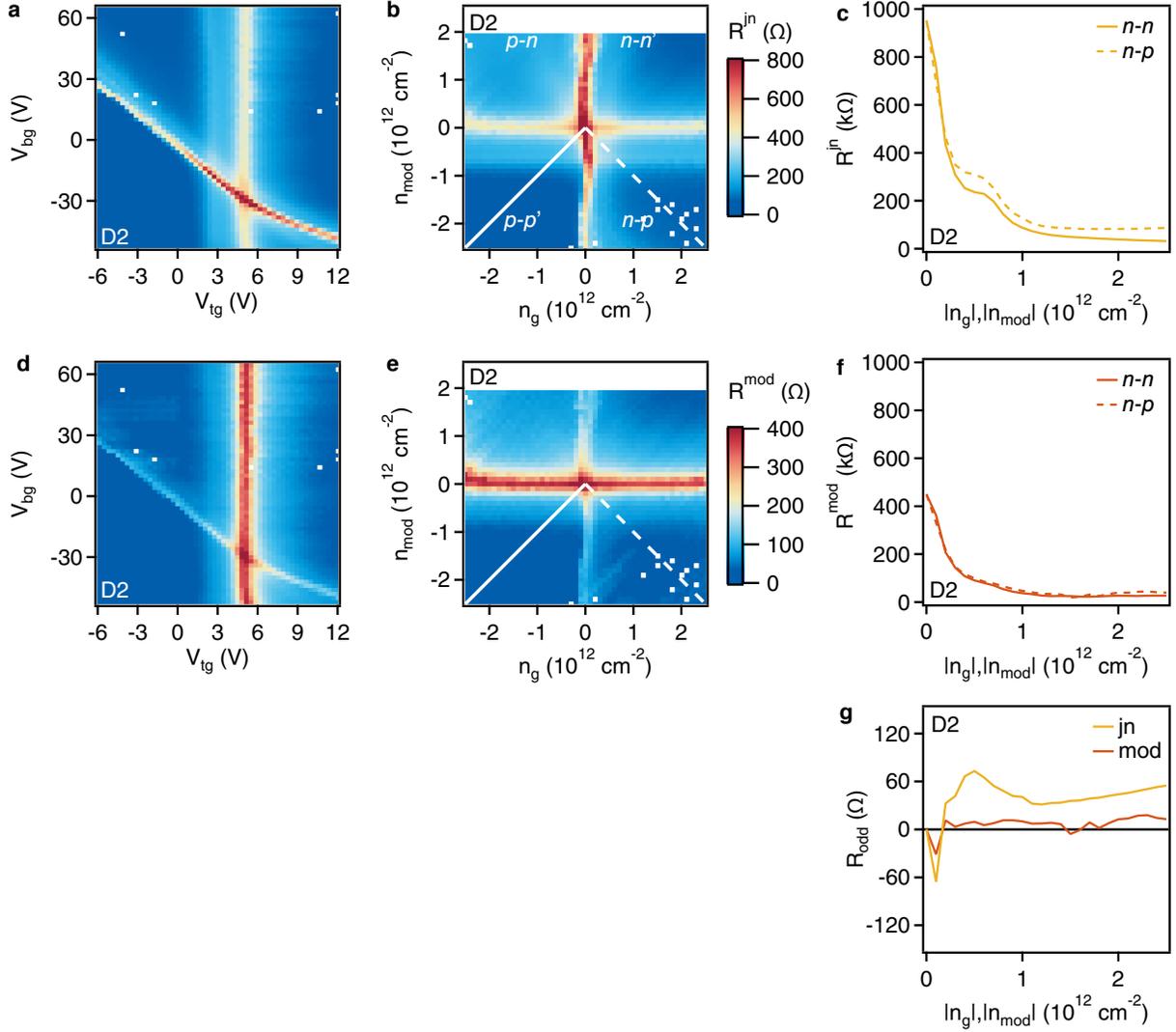

Figure S7: **Comparison of *p-p* and *n-p* resistances in device D2.** **a,d** Resistance across the junction and in the mod side as a function of top and bottom gates. **b,e,h** Resistances from **a,d,g** re-plotted as a function of g- and mod-side densities. **c,f** Linecuts along solid and dashed lines from **b,e**. **j** Differences between the curves in **c,f**.

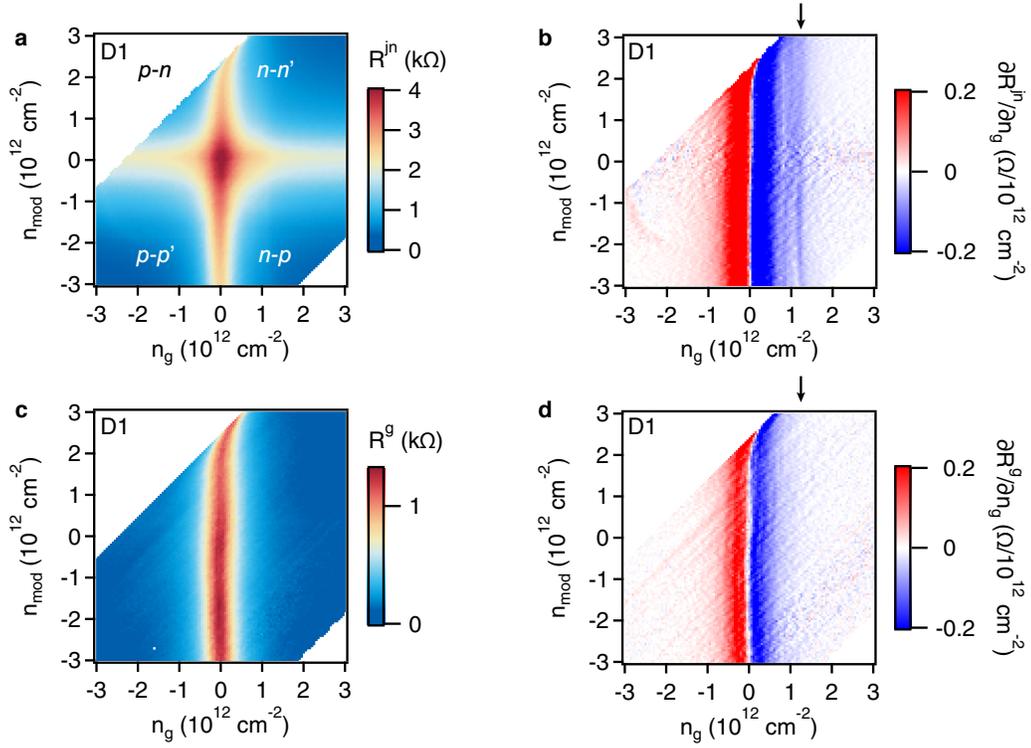

Figure S8: **g-side asymmetry in device D1.** **a** $R^{jn}$ in D1 as a function of g- and mod-side densities. **b** Derivative of **a** taken along the $n_g$-direction. **c** $R^g$ in D1 as a function of g- and mod- side densities. **d** Derivative of **c** taken along the $n_g$-direction. Panel **a** is identical to Fig. 2b of the main text, except for the lack of line cuts.

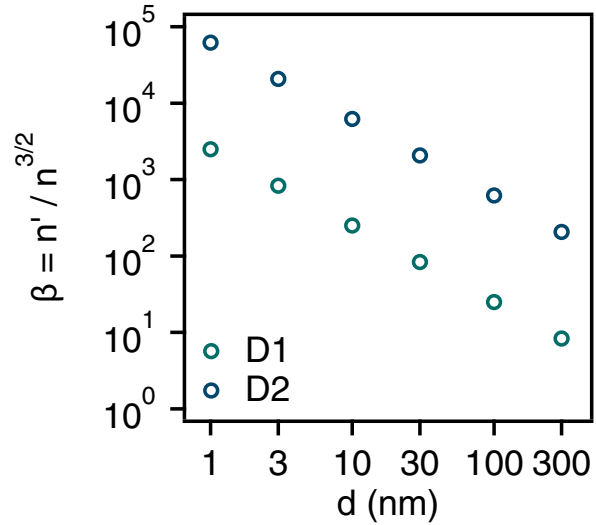

Figure S9: **Calculation of $\beta$.** $\beta$ parameter calculated for $n=10^{12}$ cm$^{-2}$ as a function of junction width $d$ for devices D1 and D2.

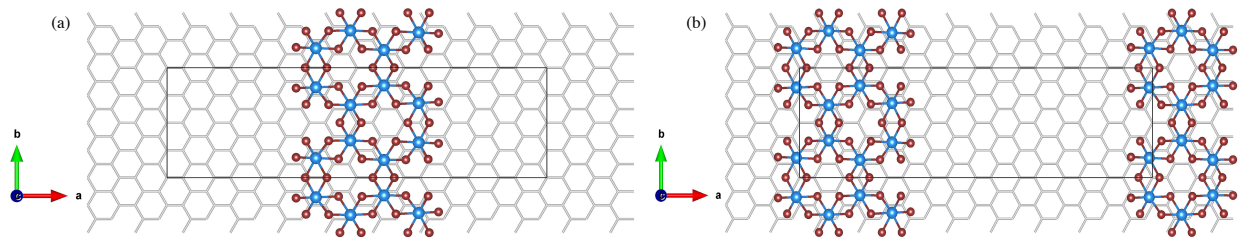

Figure S10: **Configurations considered for DFT calculations.** **a** Configuration 1. **b** Configuration 2. The gray lines indicate the graphene sheet and blue and red spheres represent the Ru and Cl atoms, respectively. The black rectangle shows the supercell in the a-b plane.

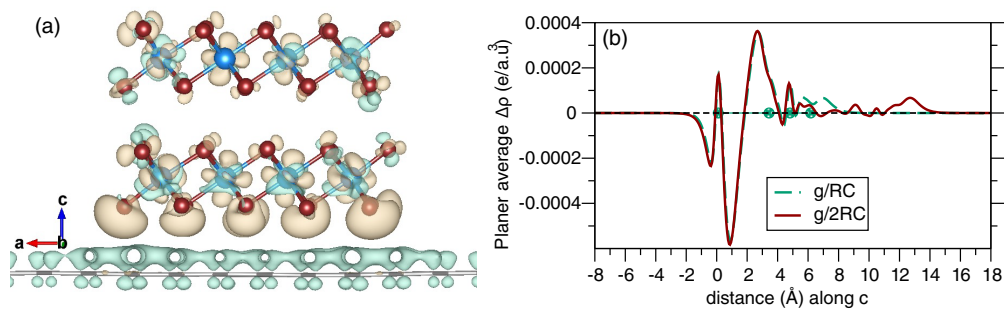

Figure S11: **Charge transfer to a bilayer $\alpha$-RuCl$_3$ above graphene.** **a** Isosurface plotted at $5\times10^{-4}$ $e$/a.u.$^3$. **b** Planar average of charge density difference (over $\hat{a}\times\hat{b}$) plotted along $\hat{c}$ direction comparing charge transfer from graphene to both a mono- and bilayer of $\alpha$-RuCl$_3$.

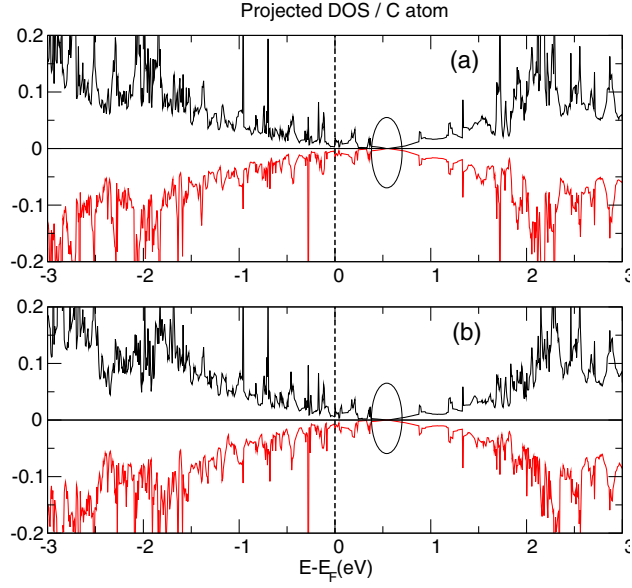

Figure S12: **Spin-polarized projected density of states (PDOS) per C atom.** (a) and (b) represent PDOS on C atoms which are far and near to the edge of $\alpha$-RuCl$_3$ ribbon, respectively. The black and red colors indicate the spin majority and minority channels, respectively. The Dirac points are circled in both the cases. As we have chosen FM configuration, the $\alpha$-RuCl$_3$ orbitals and the graphene orbitals overlap over an extended region than reported in Ref.[10]

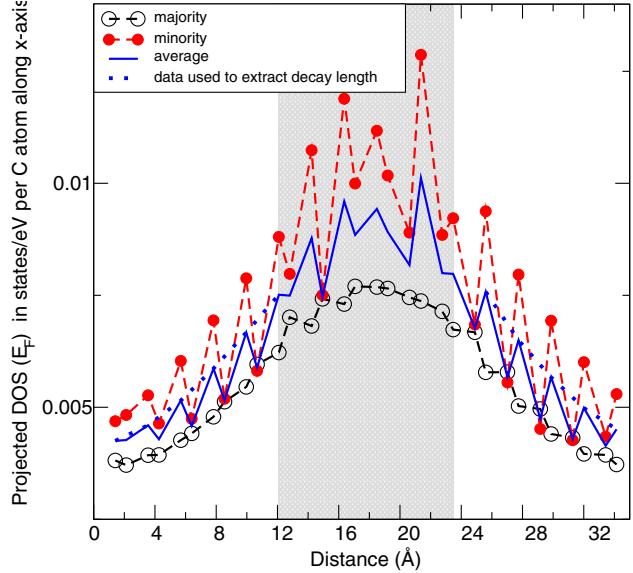

Figure S13: **Spin-polarized projected density of states (PDOS) at the Fermi energy**, per position-averaged C atom along $x$-direction. The shaded area indicates the position of the $\alpha$-RuCl$_3$ ribbon.



\end{document}